\documentclass[12pt]{article}
\usepackage[utf8]{inputenc}

\newcommand{\blind}{1}

\usepackage[style=chicago-authordate,maxcitenames=1, maxbibnames=99,backend=biber,alldates=comp, giveninits=true]{biblatex}
\AtEveryBibitem{%
  \clearfield{month}%
}

\usepackage{amsfonts,amsmath,amsthm,amssymb,color,graphicx,bm,bbm}
\usepackage{graphicx}
\usepackage{enumerate}
\usepackage{physics}
\usepackage{enumitem}
\usepackage{caption}
\usepackage{multirow}
\usepackage{enumitem}
\usepackage{booktabs}
\usepackage{mathtools}
\usepackage{setspace}
\usepackage{subcaption}
\usepackage[flushleft]{threeparttable} 
\usepackage{booktabs,caption}

\usepackage{soul}
\usepackage{bbm}
\usepackage[ruled,linesnumbered]{algorithm2e}
\definecolor{darkpowderblue}{rgb}{0.0, 0.2, 0.6}
\definecolor{goldenpoppy}{rgb}{0.99, 0.76, 0.0}
\definecolor{cardinal}{rgb}{0.77, 0.12, 0.23}
\usepackage{hyperref}
\hypersetup{
    colorlinks=true,
    linkcolor=cardinal,
    citecolor=darkpowderblue,      
    urlcolor=cyan,
}

\newcommand{\mc}[1]{\mathcal{#1}}

\newcommand{\reals}{{\mbox{\textbf{R}}}}

\DeclareMathOperator*{\argmin}{arg\,min}

\newtheorem{theorem}{Theorem}
\newtheorem*{theorem*}{Theorem}

\newtheorem{definition}{Definition}

\newtheorem{remark}{Remark}
\newtheorem{proposition}{Proposition}
\newtheorem{corollary}{Corollary}[theorem]

\newcommand{\new}[1]{\textcolor{black}{#1}}
\newcommand{\vpara}[1]{\vspace{0.01in}\noindent\textbf{#1 }}

\begin{document}

\def\spacingset#1{\renewcommand{\baselinestretch}%
{#1}\small\normalsize} \spacingset{1}


\if1\blind
{
  \title{\bf     Population-level Balance in Signed Networks}
  \author{ Weijing Tang \\
    \normalsize Department of Statistics and Data Science, Carnegie Mellon University
	\\
    and \\
    Ji Zhu \\
   \normalsize Department of Statistics, University of Michigan}
    \date{}
  \maketitle
  \vspace{-15mm}
} \fi

\if0\blind
{
  \bigskip
  \bigskip
  \bigskip
  \begin{center}
    {\LARGE\bf Population-level Balance in Signed Networks}
\end{center}
  \medskip
} \fi

\bigskip
\begin{abstract}
Statistical network models are useful for understanding the underlying formation mechanism and characteristics of complex networks. However, statistical models for \textit{signed networks} have been largely unexplored. In signed networks, there exist both positive (e.g., like, trust) and negative (e.g., dislike, distrust) edges, which are commonly seen in real-world scenarios. 
The positive and negative edges in signed networks lead to unique structural patterns, which pose challenges for statistical modeling. 
In this paper, we introduce a statistically principled latent space approach for modeling signed networks and accommodating the well-known \textit{balance theory}, i.e., ``the enemy of my enemy is my friend'' and ``the friend of my friend is my friend''. The proposed approach treats both edges and their signs as random variables, and characterizes the balance theory with a novel and natural notion of population-level balance. 
This approach guides us towards building a class of balanced inner-product models, and towards developing scalable algorithms via projected gradient descent to estimate the latent variables. 
We also establish non-asymptotic error rates for the estimates, which are further verified through simulation studies. 
In addition, we apply the proposed approach to an international relation network, which provides an informative and interpretable model-based visualization of countries during World~War~II.
\end{abstract}

\noindent%

\spacingset{1.9} 

\section{Introduction}
Networks characterize connectivity relationships between individuals of a complex system and are ubiquitous in various fields, such as social science, biology, transportation, and information technology \parencite{mark2010}. 
In a network, a node represents an individual and an edge between two nodes indicates the presence of certain interaction or relation. 
Given the unique relational information represented by networks, many statistical models have been developed to understand the underlying mechanism of the system and help explain the observed phenomenon on networks; see for example \textcite{MAL-005} for a comprehensive overview. 
One important class of statistical models is the latent variable model, where the presence/absence of an edge depends on the node latent variables. 
For example, stochastic block models use latent categorical variables to describe the block structure among nodes \parencite{JMLR:v18:16-480};
latent space models map nodes into a low-dimensional metric space while accounting for transitivity, homophily for node attributes, node heterogeneity and clustering \parencite{hoff2002latent, KRIVITSKY2009204}. 
Such latent variable models are attractive due to their interpretable structure, their nature for network visualization, and their ability for downstream network-assisted learning such as node clustering, node classification, and~network link prediction.

\begin{figure}
		\centering
    	\includegraphics[width=0.5\textwidth, height=1.7cm]{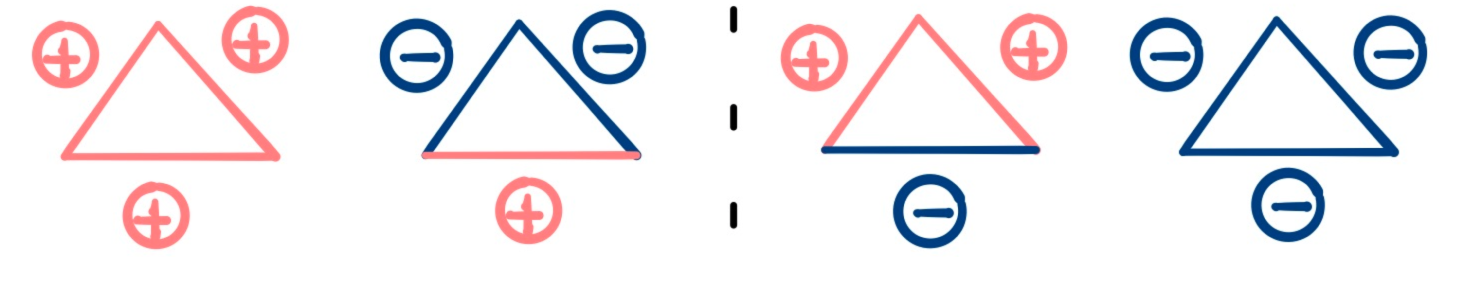}
    	\caption{Four types of triangles in signed networks, where the left two are balanced and the right two are unbalanced.}
    	\label{fig: example of triangles}
\end{figure}

Nonetheless, most statistical network models only focus on the presence/absence of edges while ignoring different types of edges, which makes them inadequate for modeling \textit{signed networks}. A signed network consists of two types of edges, positive edges and negative edges, and such polarized relationships are common in real-world networks.
For example, positive and negative edges may respectively correspond to relationships of like and dislike in social networks, collaboration and competition in trading networks, or alliance and militarized dispute in international relation networks. 
Modeling signed networks has its own unique challenges not merely due to the additional sign for each edge, but more importantly, because the presence of positive and negative edges affect each other in certain ways. 
There have been various social theories that describe the structural pattern of signed networks \parencite{guha2004propagation, leskovec2010signed}, 
an important one being the structural balance theory \parencite{harary1953notion}. 
Specifically, the balance theory describes the distribution of different types of triangles (i.e. three nodes that are connected with each other). 
A triangle in a signed network is called \textit{balanced} if the product of its three edge signs is positive; and it is called \textit{unbalanced} otherwise (see Figure~\ref{fig: example of triangles} for examples). 
The balance theory postulates that balanced triangles should be more prevalent than unbalanced triangles in signed networks, which directly coincides with the proverb, ``the enemy of my enemy is my friend'' and ``the friend of my friend is my friend''. 
Moreover, recent studies have found empirical evidence of the balance property in many real-world signed networks \parencite{leskovec2010signed, PhysRevE.99.012320, doi:10.1080/01621459.2020.1764850}.

On the other hand, there have been very few statistical models for signed networks that incorporate the balance theory into modeling. 
To the best of our knowledge, \textcite{derr2018signed} is the only recent work; specifically, it extends the configuration model \parencite{chung2006complex} to signed networks with a focus on matching not only the node degree distribution but also the sign distribution and proportion of balanced triangles.  Besides statistical models, there is a collection of work using low-rank matrix completion approaches induced by the balance theory for learning tasks such as sign prediction and clustering \parencite{hsieh2012low, JMLR:v15:chiang14a}. 
These works assume that there are underlying signed edges (not allowing for zeros) between {\it all} possible pairs of $n$ nodes, and view the network as a fixed $n\times n$ adjacency matrix with entries of $\{+1, -1\}$. 
In comparison, statistical network models can provide statistically interpretable structures and account for noise in both signs and edges by modeling network distributions that precisely quantify the randomness in the observed data.

In this paper, we propose a latent space approach to accommodate the balance theory for signed networks in a statistically principled way. 
Specifically, we introduce a novel definition of balance at the population level, which matches the balance theory in nature while viewing an observed network as the realization of a random quantity. 
For concreteness, we consider an undirected signed network with $n$ nodes denoted by a symmetric signed adjacency matrix $A$, with $A_{ij}=A_{ji}=1$ if node $i$ and node $j$ are linked by a positive edge, $A_{ij}=A_{ji}=-1$ if node $i$ and node $j$ are linked by a negative edge, and $A_{ij}=A_{ji}=0$ if there is no edge between $i$ and $j$. We assume there is no self-loop and thereby the diagonal elements of $A$ are zeros. We assume $A_{ij}$ to be random variables taking values in $\{-1,0,1\}$ and define the notion of balance at the population level as follows.
\setlength{\belowdisplayskip}{2pt} \setlength{\belowdisplayshortskip}{0pt}
\setlength{\abovedisplayskip}{2pt} \setlength{\abovedisplayshortskip}{1pt}
\begin{definition}[Population-level balanced network]
\label{def: balance}
	A network is population-level balanced~if 
	\[E(A_{ij}A_{j\ell}A_{\ell i}\big||A_{ij}A_{j\ell}A_{\ell i}|=1)>0, \text{ for any three different nodes } (i,j,\ell).\]
\end{definition}
\noindent This definition suggests the expected product of signs on any triangle to be positive but does not require all triangles to be balanced in an observed signed network. 
Furthermore, the stochastic notion in Definition~\ref{def: balance} allows us to investigate what generating mechanisms of signed networks are inherently of population-level balance. 
Specifically, we will focus on a general class of latent space models, due to aforementioned merits of latent space models. 
Rigorous descriptions are provided in Section~\ref{sec: lsm for signed network}.
The key finding is that, if there exists a partition of the latent space such that edges tend to be positive within the same subset and negative between different subsets, then the network generated from such a latent space model is population-level balanced. 

Based on this finding, we further propose a class of balanced inner-product models that directly capture the population-level balance. 
A unique difference from latent space models for unsigned networks is that we introduce an additional \textit{latent polar variable} for each node. 
In particular, when the product of latent polar variables of two nodes has a large positive value, the sign of an edge between them is more likely to be positive; for a node with the latent polar variable being zero, it has no preference on the signs when forming edges with other nodes.
We note that it is this novel introduction of latent polar variables that enables modeling signed networks with the population-level balance.

This paper is organized as follows. We introduce the latent space approach for signed networks in Section~\ref{sec: lsm for signed network}, where we also provide a sufficient condition for the population-level balance.  We present the proposed balanced inner product models in Section~\ref{sec: models}. We develop two scalable estimation methods in Section~\ref{sec: est} and establish their non-asymptotic error rates in Section~\ref{sec: thm}, which are further validated by simulation studies in Section~\ref{sec: simu}. We demonstrate the effectiveness of the proposed approach in modeling a real-world signed network for international relations in Section~\ref{sec: real data}. All proofs are given in the Supplemental~Material.

\vspace{-8mm}
\section{A Latent Space Approach for Signed Networks}
\vspace{-3mm}
\label{sec: lsm for signed network}
In this section, we propose a probabilistic generative process for undirected signed networks with $n$ nodes. 
Recall that $A\in\{1,0,-1\}^{n\times n}$ is the signed adjacency matrix. Suppose the latent space $\mathcal{U}_0$ is endowed with the probability measure $P_u$; $B(\cdot, \cdot):\mathcal{U}_0\times \mathcal{U}_0 \rightarrow (0,1) $ and $f(\cdot, \cdot):\mathcal{U}_0 \times \mathcal{U}_0 \rightarrow (-\infty, \infty)$ are two measurable symmetric functions.

\begin{definition}[A general latent space model for signed network $G(n,\mathcal{U}_0, P_u, B, f)$] 
\label{def: lsm}
For $1\leq i\leq n$, let $u_i\in \mathcal{U}_0$ be the latent vector independently sampled from the distribution $P_u$. Given the latent vectors of a pair of nodes $i$ and $j$, independently of other pairs, an edge between node $i$ and node $j$ is drawn with probability $B(u_i, u_j)$, i.e., 
\[|A_{ij}| \overset{\text{ind.}}{\sim}\text{Bernoulli}(P_{ij})\text{ \ with \ }P_{ij}= B(u_i, u_j); \] 
then for each edge (i.e. $|A_{ij}|=1$), independently of all others, it takes the positive sign with logit $f(u_i, u_j)$ and the negative sign otherwise, i.e.,
\[\text{logit}(A_{ij}=1\Big| |A_{ij}|=1)=f(u_i, u_j).\]
We write $A \sim G(n,\mathcal{U}_0, P_u, B, f)$ to denote a signed network with $n$ nodes generated from the above procedure.
\end{definition}

Note that in the network generative process in Definition~\ref{def: lsm}, the first part for generating edges covers many existing latent variable models for unsigned networks as special cases by specifying different functions $B(\cdot, \cdot)$; the second part for generating signs further models the sign distribution through specifying \new{the logit-transformed probability $f(\cdot, \cdot)$, as the sign of $f$ is the same as that of the conditional expectation of $A_{ij}$.}
Given this general class of latent space models for signed networks, next we identify the connection between the population-level balance and the key components of the model $(P_u, B, f)$. 
As we will see, this connection serves as the foremost step for incorporating the balance theory into modeling signed networks. 
The following proposition provides a sufficient condition for the symmetric function $f(\cdot, \cdot)$ such that the generated network is population-level balanced.
\vspace{-2mm}
\begin{proposition}
	\label{prop: balance for lsm} Suppose a symmetric function $f(\cdot, \cdot)$ satisfies that 
\begin{equation}
\label{eq: f balance}
	f(a,b)\cdot f(b,c)\cdot f(c,a)> 0, \text{ \ for any } a, b, c\in \mathcal{U},
\end{equation}
where $\mathcal{U}$ is a subset of $\mathcal{U}_0$ with probability $1$, i.e., $P_u(\mathcal{U})=1$. Then, a network $A\sim G(n,\mathcal{U}_0, P_u, B, f)$ \new{with a symmetric measurable function $B(\cdot, \cdot):\mathcal{U}_0\times \mathcal{U}_0 \rightarrow (0,1)$} is population-level balanced.
\end{proposition}
\vspace{-2mm}
The proof of Proposition~\ref{prop: balance for lsm} is based on the fact that $E(A_{ij}|u_i, u_j)>0$ if and only if the logit $f(u_i, u_j)>0$ when the probability that an edge appears between node $i$ and node $j$ is nonzero. The details are provided in the Supplemental Material. 
We note that though there is room for relaxation of the requirement (\ref{eq: f balance}), its simplicity provides a feasible direction for further analyses on the form of $f$. 

Since not any arbitrary symmetric function $f$ would satisfy (\ref{eq: f balance}), it is desirable to study what characteristics the function $f$ should have. 
To this end, we have established the necessary and sufficient conditions in Theorem~\ref{thm: balance} for the function $f$ to satisfy (\ref{eq: f balance}).
\begin{theorem}
	\label{thm: balance} For a symmetric function $f(\cdot, \cdot):\mathcal{U}_0 \times \mathcal{U}_0 \rightarrow (-\infty, \infty)$,  $f(a,b)\cdot f(b,c)\cdot f(c,a)> 0$ holds for any $a,b,c\in\mathcal{U}$, where $\mathcal{U}\subset \mathcal{U}_0$ with $P_u(\mathcal{U})=1$, 
	if and only if
	\begin{enumerate}[label=(\roman*)]
		\item the function $f$ is positive on $\mathcal{U}\times \mathcal{U}$, i.e., $f(a,b)>0$ for any $a,b\in \mathcal{U}$; or
		\item there exists two nonempty subsets $S$ and $T$, with $S\cup T = \mathcal{U}$ and $S \cap T=\varnothing$, such that $sign(f(a,b))= \mathbbm{1}(a\in S)\cdot \mathbbm{1}(b\in S)$ for any $a,b\in \mathcal{U}$, where $\mathbbm{1}(event)$ is not the usual indicator function, but rather equals $1$ if the event holds and $-1$ otherwise.
	\end{enumerate}
\end{theorem}

Theorem~\ref{thm: balance} implies that, for a function $f$ to satisfy (\ref{eq: f balance}), if it is not always positive, then $\mathcal{U}$ can be divided into two nonempty disjoint subsets such that the function $f$ is positive when the two arguments belong to the same subset and negative otherwise. 
On the other hand, if the function $f$ is always positive, it corresponds to a trivial case in which the expected signs between all pairs of nodes are positive. 
Note that when $\mathcal{U}$ is discrete and finite, Theorem~\ref{thm: balance} is a direct result of \textcite{harary1953notion}, while our theorem can be applied to more general latent spaces.


Next, we further illustrate the implication of Theorem~\ref{thm: balance} in choosing the function $f$  to describe the population-level balance by taking commonly used latent spaces as examples. 

\begin{figure}
\begin{subfigure}{.5\textwidth}
  \centering
  \includegraphics[width=.7\linewidth]{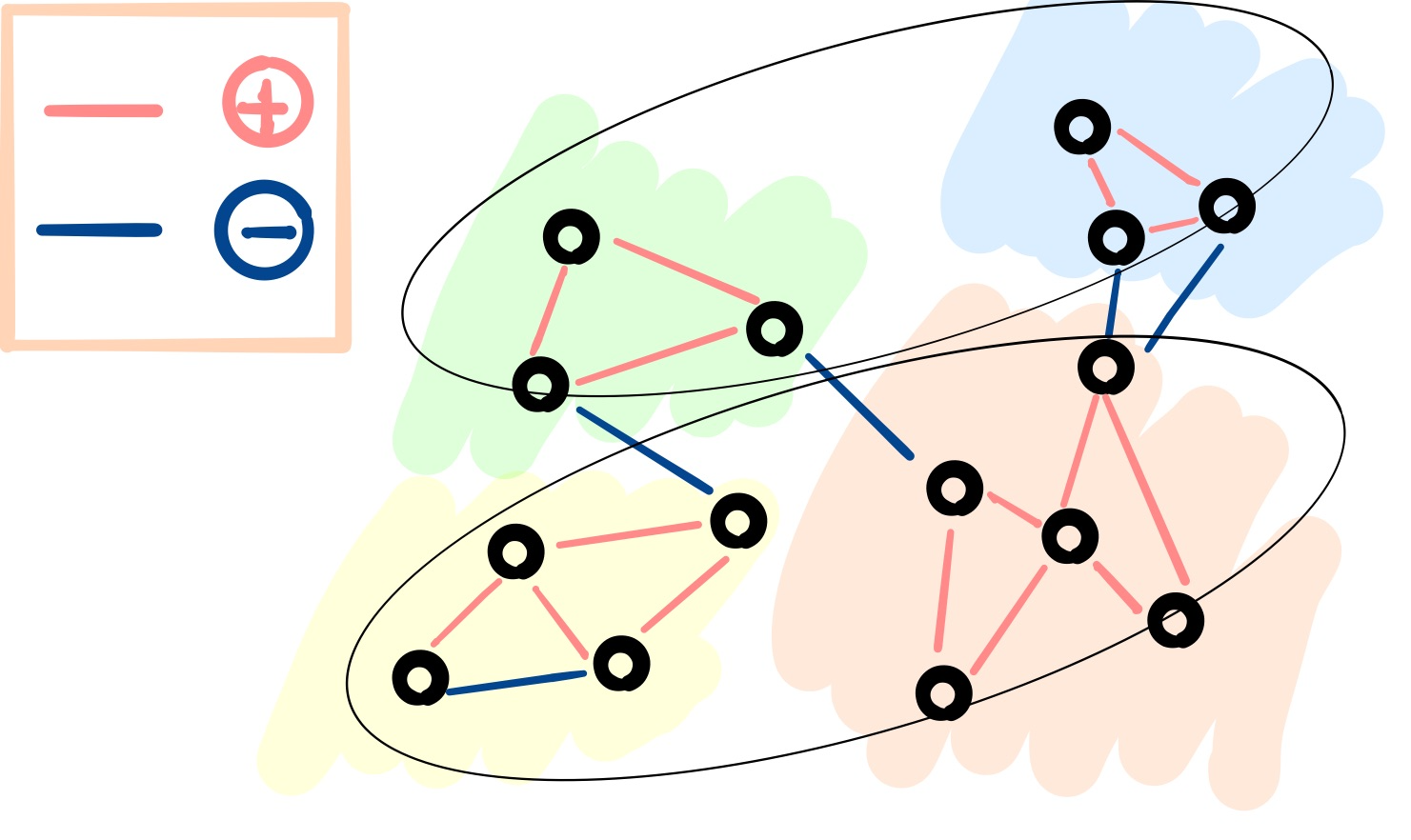}
\end{subfigure}%
\begin{subfigure}{.5\textwidth}
  \centering
  \includegraphics[width=.61\linewidth]{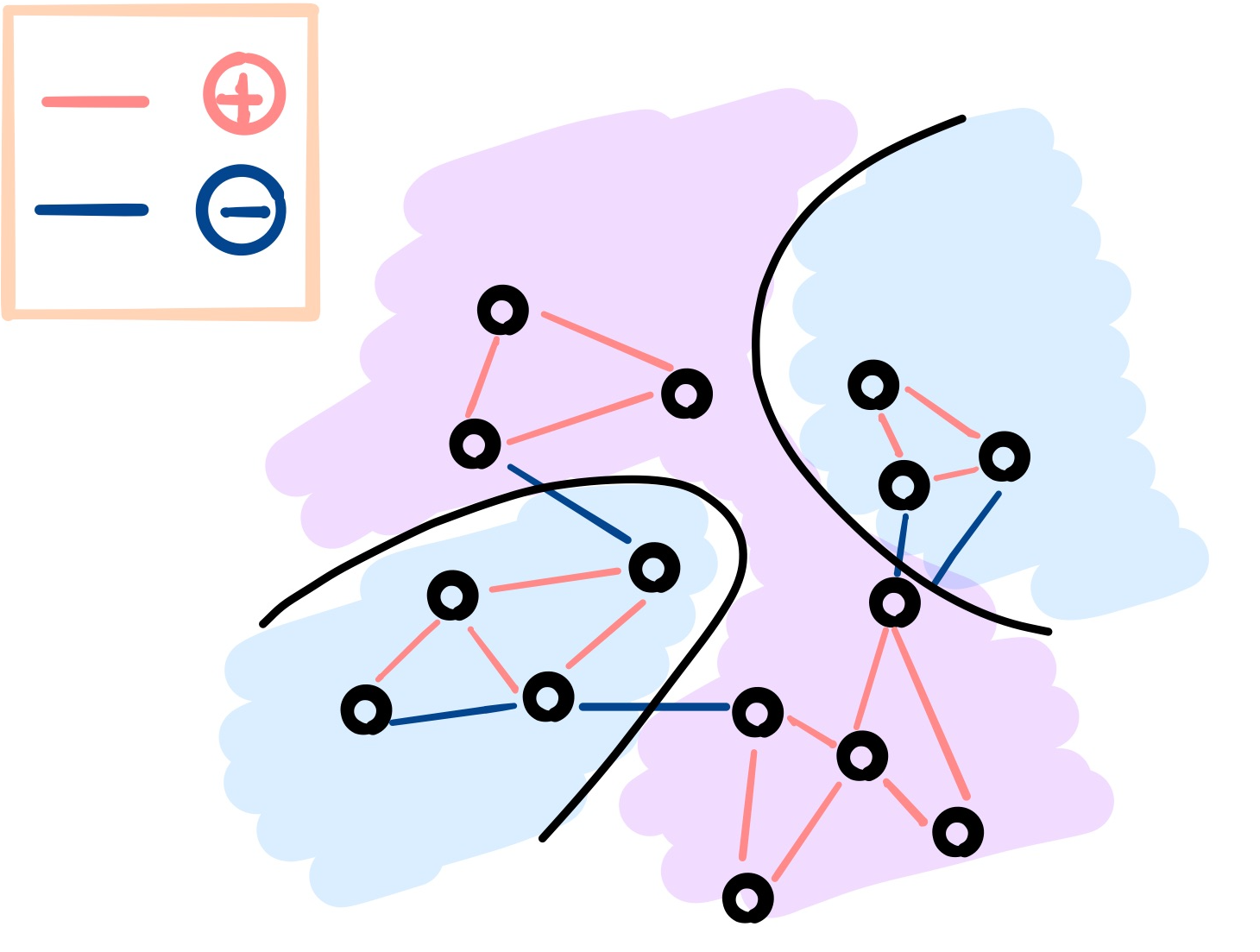}
\end{subfigure}
\caption{Illustration of the latent space partition in Theorem~\ref{thm: balance}. Left: $\mathcal{U}_0$ is a finite set, where each color corresponds to a possible state in $\mathcal{U}_0$ and the two ellipses correspond to the partition. 
Right: $\mathcal{U}_0$ is a Euclidean space, where the two colors correspond to the partition. }
\label{fig:examples of theorem 1}
\end{figure}

\vpara{Example 1} The latent space $\mathcal{U}_0$ can be a finite set as in stochastic block models \parencite{JMLR:v18:16-480}. 
Let $\mathcal{U}_0=\{1, \cdots, K\}$ and $u_i$ denotes the community that node $i$ belongs to. 
Theorem~\ref{thm: balance} implies that the $K$ communities can be further combined into two groups and edges tend to be positive within the same group and negative between different groups, as shown in the left side of Figure~\ref{fig:examples of theorem 1}. We provide a rigorous description for the above result in the following corollary.

\vspace{-2mm}
\begin{corollary}
\label{corol: f for sbm}
	For a finite set $\mathcal{U}_0=\{1, \dots, K\}$, the symmetric function $f(\cdot, \cdot):\mathcal{U}_0 \times \mathcal{U}_0 \rightarrow (-\infty, \infty)$ satisfies that  
	$f(a,b)\cdot f(b,c)\cdot f(c,a)> 0 \text{ for any } a,b,c \in \mathcal{U}_0,$ 
	if and only if there exists a grouping function $g:\{1, \dots, K\}\rightarrow \{-1,1\}$ and some constants $q_{ab}=q_{ba}>0$ such that $f(a,b)=q_{ab}\cdot g(a)\cdot g(b)$ holds for $1\leq a,b\leq K$.
\end{corollary}

\noindent Note that here the grouping function $g$ identifies two antagonistic groups in the signed network, where nodes from different groups tend to ``dislike'' each other.

\vpara{Example 2} The latent space can also be a Euclidean space as in the latent distance model and the latent projection model \parencite{hoff2002latent}. 
The following proposition provides an important class of continuous symmetric functions for which the requirement (\ref{eq: f balance}) is satisfied. 
\begin{proposition}
\label{prop: example of f}
	For the Euclidean space $\mathcal{U}_0=\reals^k$, the requirement (\ref{eq: f balance}) holds if  $f(a, b)=\phi(a)\phi(b)$, where $\phi(\cdot)$ is a real-valued continuous function and $P_u(u: \phi(u)\neq 0)=1$.
\end{proposition}

\noindent From the mathematical perspective, Proposition \ref{prop: example of f} is an obvious result based on the inequality in (\ref{eq: f balance}). However, in combination with Theorem~\ref{thm: balance}, it leads us to a useful and interesting interpretation for the function $\phi$. 
Specifically, $\phi (\cdot)$ can be viewed as the logit (or score) of any 
binary ``classifier'' that separates $\mathcal{U}_0$ into two disjoint regions; the function $f$ is positive if the two arguments are classified into the same region and negative otherwise.
As shown in the right panel of Figure~\ref{fig:examples of theorem 1}, the boundary of the classifier tries to cut as many negative edges as possible while retaining most positive edges within the same region. 
with some $w\in\reals^k$ and $\gamma \in \reals$.

\begin{remark}
\new{
   For the special case $f(a,b)=a^\top b$, \textcite{hoff2005bilinear} discussed the use of the inner product of two latent variables to capture various third-order dependence patterns among residuals in the context of regression, including the balance theory. Specifically, when the dimension $k$ is one, \textcite{hoff2005bilinear} observed that all triads among residuals are balanced, which allows for clustering nodes into two groups based on signs of latent variables. This observation aligns with our findings on the separation of the latent space. Nonetheless, the balance concept studied in \textcite{hoff2005bilinear} is at the triangle level and does not extend to the entire network structure. In this work, we have introduced a rigorous stochastic definition of population-level balance for signed networks. Furthermore, the condition in Proposition~1 for population-level balance is more general and includes the inner product~as~an~example.}
\end{remark}


\begin{remark}
	Moreover, our findings in this section can be generalized beyond triangles. In the Supplemental Material, we generalize the notion of population-level balance and extend Proposition~\ref{prop: balance for lsm} to any $\ell$-loops as well as general weighted networks. 
  \new{In addition, we extend the infinitely exchangeable graphon model for signed networks and expand Theorem~\ref{thm: balance} accordingly.} 
\end{remark}

\section{Balanced Inner-product Models}
\label{sec: models}
Motivated by the key finding in Theorem~\ref{thm: balance}, we propose two inner-product models for signed networks that fall within the general class of latent space models in Definition~\ref{def: lsm}.
Both models are inherently of population-level balance. 
We first present the separate inner-product model and then introduce the joint inner-product model by adding an additional structural assumption. 
We will demonstrate the usefulness of this structural assumption in estimating the latent variables both theoretically (if it is correctly specified) and empirically in subsequent sections.

\subsection{Separate inner-product model}
We assume that for any $1 \le i<j \le n$, we have
\begin{equation}
\label{eq: model step 1}
	|A_{ij}|=|A_{ji}|\overset{\text{ind.}}{\sim}\text{Bernoulli}(P_{ij}),\text{ \ with \ logit}(P_{ij})=\Theta_{ij}= \alpha_i+\alpha_j + z_i^\top z_j, 
\end{equation}
where $\alpha_i\in \reals$ and $z_i\in\reals^k$ (for $i=1, \ldots, n$) are latent variables. Further, independently of all others, we also assume
\begin{equation}
\label{eq: model step 2}
	\text{logit}(A_{ij}=1\Big| |A_{ij}|=1)=\eta_{ij}=v_iv_j,
\end{equation}
where $v_i \in\reals$ for $i=1,\ldots, n$ are also latent variables.

The proposed separate inner-product model has the capacity to capture various commonly observed characteristics of signed networks. 
Specifically, the parameter $\alpha_i$ enables modeling node degree heterogeneity, of which a larger value  leads to higher probability of connecting with other nodes given other parameters fixed. Thus, we call $\{\alpha_i\}_{i=1}^n$ {\it degree heterogeneity parameters}. 
Next, the inner-product formation between the {\it latent position vectors} $z_i$ and $z_j$ inherently models transitivity, i.e., nodes with common neighbors (regardless of friend or enemy) are more likely to be linked. Because the closer the latent position vectors of two nodes are in the latent space, the higher inner product it is and more likely to connect with each other.
Finally, the parameters $\{v_i\}_{i=1}^n$ model the distribution of signs through their product, which satisfies the sufficient condition (\ref{eq: f balance}) for the population-level balance. 
In particular, an edge between two nodes tends to have a positive sign when their latent variables $v_i$ and $v_j$ have the same sign and a negative sign otherwise. Moreover, the magnitude of $v_i$ controls the discrepancy level between positive and negative signs. Therefore, we name them as {\it latent polar variables}. When all latent polar variables are zeros, negative and positive signs are exchangeable. 

Following \textcite{ma2020universal}, which is for unsigned networks, we impose no distributional assumptions (prior) on the latent variables $(\alpha_i, z_i, v_i)$ for the sake of modeling flexibility and estimation scalability, in comparison to treating them as random and using Bayesian estimation approaches as in existing works \parencite{KRIVITSKY2009204}.

For presentation simplicity, we rewrite the model in matrix form. Specifically, we have  
$\Theta=\alpha 1_n^\top+1_n\alpha^\top+ZZ^\top, \ \ \eta=vv^\top$,
where $\alpha=(\alpha_1, \cdots, \alpha_n)$, $1_n$ is the all one vector in $\reals^n$, $Z=(z_1, \cdots, z_n)^\top\in \reals^{n\times k}$, and $v=(v_1, \cdots, v_n)^\top \in \reals^n$. 

\vpara{Identifiability} To ensure identifiability of parameters $(\alpha, Z, v)$, we provide  additional constraints in Proposition~\ref{prop: iden for v}.  Given centered latent position variables, that is $J_nZ=Z$, where~$J_n=I_n-\frac{1}{n} 1_n 1_n^\top$, 
the parameters are identifiable up to an orthogonal transformation and a sign flipping.

\begin{proposition}
	\label{prop: iden for v}
	Suppose two sets of parameters $(\alpha, Z, v)$ and $(\bar{\alpha}, \bar{Z}, \bar{v})$ satisfy that A1) $J_nZ=Z$ and $J_n\bar{Z}=\bar{Z}$; A2) $Z\in\reals^{n\times k}$ is of full rank. Then, they specify the same network distribution through (\ref{eq: model step 1}) and (\ref{eq: model step 2}) 
	if and only if there exist an orthogonal matrix $O\in \reals^{k\times k}$ with $O^\top O=OO^\top=I_k$ and $\kappa\in\{-1,1\}$ such that $\alpha=\bar{\alpha},Z = \bar{Z}O, v=\kappa \bar{v}$.
\end{proposition}

\subsection{Joint inner-product model}
Based on the above separate inner-product model, we further consider the dependency of the latent polar variable $v_i$ on the latent position variable $z_i$. 
The idea of introducing their relationship originates naturally from Proposition~\ref{prop: example of f}, where we view the latent polar variable $v_i$ as a function of the latent position variable $z_i$, i.e., $v_i=\phi(z_i)$ with some link function $\phi$. 
Modeling such a link function $\phi$ can provide more structural information of the network. 
On the other hand, there are flexibilities in choosing the family of link function $\phi$, which would lead to different shapes of the latent space partition derived by $\phi$. 
For the scope of this paper, we assume $\phi$ is a linear function in $z_i$ in the joint inner-product model, i.e., $v_i=w^\top z_i+\gamma$ with $w\in \reals^k$ and $\gamma\in \reals$, and discuss other nonlinear alternatives in Remark~\ref{remark: nonlinear examples}. 
Specifically, the joint inner-product model is given by (\ref{eq: model step 1}) and replaces (\ref{eq: model step 2}) with
\begin{equation}
\label{eq: model step 3}
	\text{logit}(A_{ij}=1\Big| |A_{ij}|=1)=\eta_{ij}=(w^\top z_i+\gamma)(w^\top z_j+\gamma).
\end{equation}
In particular, the hyperplane $\{z\in \reals^k: w^\top z+\gamma=0\}$ separates the latent space into two regions. A pair of nodes tend to have a positive edge when their latent positions are located on the same side of the hyperplane and have a negative edge when their latent positions are located on different sides of the hyperplane. If $w=0$ and $\gamma\neq 0$, the sign of each edge has a homogeneous logit $\gamma^2$ to be positive.

\vpara{Identifiability} For the joint inner-product model, the identifiability condition for parameters $(\alpha, Z, w,\gamma)$ is established correspondingly in Proposition~\ref{prop: iden for linear}.

\begin{proposition}
	\label{prop: iden for linear}
	Suppose two sets of parameters $(\alpha, Z, w, \gamma)$ and $(\bar{\alpha}, \bar{Z}, \bar{w}, \bar{\gamma})$ satisfy that A1) $J_nZ=Z$ and $J_n\bar{Z}=\bar{Z}$; A2) $Z\in\reals^{n\times k}$ is of full rank. Then, they specify the same network distribution through (\ref{eq: model step 1}) and (\ref{eq: model step 3})
	if and only if there exist an orthogonal matrix $O\in \reals^{k\times k}$ with $O^\top O=OO^\top=I_k$ and $\kappa\in\{-1,1\}$ such that $\alpha=\bar{\alpha},Z = \bar{Z}O, w=\kappa O^\top \bar{w}, \gamma=\kappa \bar{\gamma}$.
\end{proposition}

\begin{remark}
\label{remark: nonlinear examples}
Though we use a linear link function in the joint inner-product model (\ref{eq: model step 3}), more flexible nonlinear functions can be considered. 
For example, we may assume $\phi$ belongs to a reproducing kernel Hilbert space (RKHS) $\mathcal{H}$ associated with an inner product $\langle \cdot, \cdot \rangle_{\mathcal{H}}$ under which $\mathcal{H}$ is complete.   
There is a positive semidefinite kernel function $\mathbb{K}(\cdot, \cdot): \reals^k\times \reals^k \rightarrow \reals_{+}$ such that $\phi(z_i)=\langle \phi, \mathbb{K}(\cdot, z_i)\rangle_{\mathcal{H}}$. 
Multiple choices of RKHS are available for practical use, including those with polynomial kernel, Gaussian kernel, and Laplacian kernel \parencite{scholkopf2018learning}.
\end{remark}

\vspace{-8mm}
\section{Model Estimation}
\label{sec: est}
\vspace{-3mm}
In this section, we develop two methods for fitting the proposed models (\ref{eq: model step 1})-(\ref{eq: model step 3}). 
Both methods minimize the negative log-likelihood function of the balanced inner-product models through projected gradient descent.

Under balanced inner-product models, the negative log-likelihood function consists of two parts.
The first part is derived from the probability of forming edges:
\[\mathcal{L}_e(\alpha, Z) 
= \sum_{i<j} \Big\{ |A_{ij}|\Theta_{ij}+\log(1-\sigma(\Theta_{ij}))\Big\},\]
where $\Theta=\alpha 1_n ^\top+1_n\alpha^\top+ZZ^\top$, and $\sigma(x)=1/(1+\exp(-x))$ is the sigmoid function, which is the inverse of the logit function.  The second part is derived from the probability of assigning signs:
\begin{align*}
	\mathcal{L}_s(v) 
	& = \sum_{i<j}\Big\{|A_{ij}|\frac{1+A_{ij}}{2}\eta_{ij}+|A_{ij}|\log(1-\sigma(\eta_{ij}))\Big\},
\end{align*}
where $\eta=vv^\top$, and when under the joint inner-product model, we further have $v=Zw+\gamma 1_n$, or equivalently, $v$ belongs to the column space of $(1_n, Z)$.

The first method estimates parameters $(\alpha, Z)$ and $v$ separately by minimizing $\mathcal{L}_e(\alpha, Z)$ and $\mathcal{L}_s(v)$ respectively. Hence we name it the separate estimation method. 
Note that the separate estimation method does not depend on a specific relationship between the latent polar variables $v$ and the latent position vectors $Z$. Therefore, the separate estimation method can always be applied regardless of the underlying link function $\phi$.
Alternatively, we also propose a joint estimation method tailored for the joint inner-product model, which exploits the structural assumption for more accurate estimation. Specifically, we jointly estimate parameters $(\alpha, Z, v)$ by minimizing a weighted sum of $\mathcal{L}_e(\alpha, Z)$ and $\mathcal{L}_s(v)$, while constraining $v$ to be in the column space of $(1_n, Z)$. 

\vpara{Notation} Before presenting the algorithm details, we first introduce the following general notations to be used hereafter. For any $X\in\reals^{d_1 \times d_2}$, $X_{i*}$ and $X_{*j}$ denote the $i$-th row and $j$-th column of matrix $X$ respectively, and for any function $\omega(\cdot)$, $\omega(X)$ represents applying the function $\omega(\cdot)$ element-wisely to $X$, that is $\omega(X)\in \reals^{d_1 \times d_2}$ and $[\omega(X)]_{ij}=\omega(X_{ij})$. We use $\circ$ to denote the Hadamard product, that is, for any two matrices $X, Y\in\reals^{d_1 \times d_2}$, $X\circ Y \in \reals^{d_1 \times d_2}$ and $[X\circ Y]_{ij}=X_{ij}Y_{ij}$. Moreover, we use $\|X\|_F$, $\|X\|_{op}$, $\|X\|_{*}$, and $\|X\|_{\max}$ to denote the Frobenius norm, the operator norm, the nuclear norm, and the max norm of a matrix respectively. We use $col(X)$ to denote the column space of $X$. For a vector $x\in \reals^d$, we use $\|x\|$ to denote the Euclidean~norm.

\vspace{-7mm}
\subsection{Separate estimation method} 
\vspace{-3mm}
First, to estimate parameters $(\alpha,Z)$, we solve the non-convex optimization problem below:
\begin{equation}
	\label{eq: obj for separate z}
	\min_{\alpha\in \reals, Z\in \reals^{n\times k}} -\sum_{i,j}  \Big\{|A_{ij}|\Theta_{ij}+\log(1-\sigma(\Theta_{ij}))\Big\},
\end{equation}
subject to $\Theta=\alpha 1_n ^\top+1_n\alpha^\top+ZZ^\top \text{ and } Z=J_n Z$. 
In particular, the signed adjacency matrix enters the objective function through its absolute value, which leads to the same optimization problem studied in \textcite{ma2020universal} when there is no edge covariate.
Here we adopt the projected gradient descent algorithm along with the initialization method proposed in \textcite{ma2020universal} because of their theoretical guarantee and scalability to large networks. We provide the detailed description of the method in Algorithm~\textcolor{cardinal}{S1} and the initialization algorithm in the Supplemental Material. 

Next, to estimate the latent polar variables $v$, we solve another non-convex optimization problem, i.e.,
\begin{equation}
	\label{eq: obj separate v}
	\min_{v\in \reals^{n}} -\sum_{i,j}  |A_{ij}| \Big\{\frac{1+A_{ij}}{2}\eta_{ij}+\log(1-\sigma(\eta_{ij}))\Big\} \text{ subject to } \eta=vv^\top.
\end{equation}
Similarly, we develop a fast gradient descent algorithm, which is summarized in Algorithm~\textcolor{cardinal}{S2}. We also use an initialization algorithm based on the universal singular value thresholding proposed by \textcite{chatterjee2015} (see the Supplemental Material).

\begin{remark}
	We note that, although we use gradient descent algorithms for estimating both $(\alpha, Z)$ and $v$, the subtle difference in their objectives makes the theory in \textcite{ma2020universal} not directly applicable to (\ref{eq: obj separate v}). Specifically, unlike the objective in (\ref{eq: obj for separate z}), not all elements of the signed adjacency matrix contribute to the objective in (\ref{eq: obj separate v}). 
Instead, only nonzero entries, i.e., $\{(i,j): |A_{ij}|=1\}$, are used for inferring the latent polar variables through (\ref{eq: obj separate v}). In this case, one key step in building the improvement in errors of iterates in \textcite{ma2020universal} no longer holds. Therefore, we establish a new error bound for Algorithm~\textcolor{cardinal}{S2}~(see Section~\ref{sec: thm}).
\end{remark}

\begin{remark}
We also note that our optimization problem in (\ref{eq: obj separate v}) is closely related to the line of research on low-rank matrix estimation. See \textcite{koltchinskii2011, doi:10.1137/110848074, davenport20141, chen2015fast, zheng2016convergence, wang2017unified} for a sample of references. 
In particular, (\ref{eq: obj separate v}) can be viewed as a one-bit matrix completion problem, where we observe a random subset of binary entries generated from a distribution determined by a low-rank matrix. 
To solve this problem, \textcite{davenport20141} considered a convex relaxation that replaces the low-rank constraint by the nuclear norm penalization. Though it becomes a convex optimization problem, in general, solving such a nuclear-norm penalized optimization problem requires singular value decomposition at each iteration, which is computationally expensive for large matrices. Alternatively, gradient descent algorithms have been used for improving the computational efficiency. \textcite{chen2015fast} and \textcite{wang2017unified} have established convergence guarantees and statistical errors for the gradient descent algorithms in application to low-rank matrix estimation problems, which particularly cover the one-bit matrix completion problem. However, theories in aforementioned works are~based~on~the~uniform random sampling assumption, i.e., each entry of the matrix is observed independently with a uniform probability $p$, while in our case, entries are observed with different probabilities $P_{ij}$. Thus, our theoretical analysis of the proposed gradient descent algorithm in Section~\ref{sec: thm} provides new results for one-bit matrix completion under non-uniform random~sampling.
\end{remark}

\vspace{-7mm}
\subsection{Joint estimation method}
\vspace{-3mm}
Under the joint inner-product model (\ref{eq: model step 2})-(\ref{eq: model step 3}), we propose to jointly estimate parameters $(\alpha, Z, v)$ by re-parameterizing $v=Zw+\gamma 1_n$ with $w \in \reals^k$ and $\gamma \in \reals$. 
By introducing a hyperparameter $\lambda$, we minimize the following  weighted negative log-likelihood,
\[ \mathcal{L}_\lambda(\alpha, Z, w, \gamma)=- \sum_{i,j} \Big\{(1-\lambda)  \big[|A_{ij}|\Theta_{ij}+\log(1-\sigma(\Theta_{ij}))\big]+\lambda |A_{ij}| \big[\frac{1+A_{ij}}{2}\eta_{ij}+\log(1-\sigma(\eta_{ij}))\big]\Big\},\]
subject to  $\Theta=\alpha 1_n ^\top+1_n\alpha^\top+ZZ^\top$, $Z=JZ$, and $\eta=(Zw+\gamma1_n)(Zw+\gamma1_n)^\top$. 

Here $\lambda$ controls the weight of relative information from the edge formation and the sign assignment respectively. In particular, when $\lambda=0$, no information from the edge signs is used and the joint estimation reduces to the separate estimation for $(\alpha, Z)$ in (\ref{eq: obj for separate z}). Later in Section~\ref{subs: theory for joint}, we will theoretically show that, under certain conditions, any positive $\lambda$ below some threshold  yields more accurate estimation of latent position variables $Z$ than the separate estimation (i.e., $\lambda=0$), but the magnitude of the improvement depends on the choice of $\lambda$. In principle, we can select $\lambda$ in a data-driven manner by performing cross-validation on the observed signed adjacency matrix, where we randomly mask a subset of entries, fit the joint inner-product model by using the remaining entries, repeat the process multiple times, and then select $\lambda$ from a candidate set with the best average prediction performance on the holdout entries. In practice, we find simply setting $\lambda=1/2$ also works generally well, in which case the solution becomes the usual maximum likelihood estimator.

To solve the above constrained minimization problem, we develop a projected gradient descent algorithm, whose details are given in Algorithm~\textcolor{cardinal}{S5} in the Supplemental Material. We initialize the algorithm by $(\alpha_0, Z_0)=(\hat{\alpha}, \hat{Z})$ obtained from Algorithms~\textcolor{cardinal}{S1} and $(w_0, \gamma_0)=\argmin_{w\in \reals^k, \gamma \in \reals} \mc{L}_\lambda(\alpha_0, Z_0, w, \gamma)$.

\section{Theoretical Results}
\label{sec: thm}
In this section, we establish high probability error bounds for the proposed two estimation methods. 
Note for the separate estimation method, the error bound for estimating latent position vectors~$Z$ under model~(\ref{eq: model step 1}) and that for estimating latent polar variables~$v$ under model~(\ref{eq: model step 2}) are derived separately. 
Thus, the separate estimation method is robust in the sense that, when one of models (\ref{eq: model step 1}) and (\ref{eq: model step 2}) is mis-specified, our theoretical results still hold for the other. 
On the other hand, for the joint estimation method that utilizes the relationship between $v$ and $Z$, we further discuss how incorporating their dependency can help reduce the estimation error of latent variables under the joint inner-product model~(\ref{eq: model step 3}). 

\subsection{Results for the separate estimation method}
We present theoretical guarantees of Algorithms~\textcolor{cardinal}{S1} and~\textcolor{cardinal}{S2} under the separate inner-product model~(\ref{eq: model step 1}) and model~(\ref{eq: model step 2}) respectively. 
Note that the error bound for the outputs of Algorithm~\textcolor{cardinal}{S1} is a straightforward result of \textcite[Theorem 9]{ma2020universal} when there is no edge covariate; we adjust it in Proposition~\ref{prop: separate est z error rate}  for presentation coherence. 
Nonetheless, their theory cannot be directly applied to the setting of Algorithm~\textcolor{cardinal}{S2}, because only nonzero entries of the signed adjacency matrix are included in the objective (\ref{eq: obj separate v}), which breaks an important step towards establishing the estimation improvements for successive iterations in their proof. Thus, our established error bound for the outputs of Algorithm~\textcolor{cardinal}{S2} is a new result for a more general setting, where entries are observed with non-uniform~probabilities.

We describe error bounds for the outputs of Algorithms~\textcolor{cardinal}{S1} and~\textcolor{cardinal}{S2} with details below. 
We firstly define the parameter spaces as 
\begin{align}
	\mathcal{F}_{\theta}(n, k, M_1, M_2)=& \Big\{\alpha \in \reals^n, Z \in \reals^{n\times k}, \Theta \in \reals^{n \times n}  \ |\ \Theta=\alpha 1_n ^\top+1_n\alpha^\top+ZZ^\top, J_nZ=Z, \nonumber \\
	& \ \ \ \ \ \ \ \   \max_{1\leq i\leq n} \|Z_{i*}\|^2, 2\|\alpha\|_{\max} \leq \frac{M_1}{2}, \max_{1\leq i\neq j\leq n}\Theta_{ij}\leq -M_2 \Big\} \label{eq: para space 1}
\end{align}
and 
\begin{align}
	\mathcal{F}_{\eta}(n, M_3)= \Big\{ v\in \reals^n, \eta \in \reals^{n \times n}  \  | \ \eta=vv^\top,  \|v\|_{\max}^2\leq M_3 \Big\}, \label{eq: para space 2}
\end{align}
\new{where $M_i>0$ for $i=1,2,3$.} 
We allow $k$, $M_1$, $M_2$, and $M_3$ in (\ref{eq: para space 1})-(\ref{eq: para space 2}) to change with the network size $n$ similarly as in \textcite{ma2020universal}. Note that, given the inequalities in (\ref{eq: para space 1}), it is straightforward to see that, for any $\Theta \in \mathcal{F}_{\theta}(n, k, M_1, M_2)$, we have $-M_1 \le \Theta_{ij} \le -M_2$ for $1\le i\neq j \le n$. Therefore, $M_2$, as the upper bound of logit-transformed probabilities of observing edges, controls the network sparsity, of which a larger value leads to a sparser network. 
The true parameters are denoted by $(\alpha^*, Z^*, v^*)$, $\Theta^*=\alpha^* 1_n ^\top+1_n{\alpha^*}^\top+Z^*{Z^*}^\top$, and $\eta^*=v^*{v^*}^\top$.

\vpara{Error bound for Algorithm~\textcolor{cardinal}{S1}} 
Let $(\alpha_t, Z_t)$ be the updated parameters at the $t$-th iteration in Algorithm~\textcolor{cardinal}{S1} and $\Theta_t=\alpha_t 1_n ^\top+1_n{\alpha_t}^\top+Z_t{Z_t}^\top$. 
Since the latent position vectors $Z\in \reals^{n\times k}$ are identifiable up to an orthogonal transformation, we define the distance between two latent matrices $Z_1$ and $Z_2$ as 
$dist(Z_1, Z_2) = \min_{O\in O(k)}\|Z_1-Z_2O\|_F$, 
where $O(k)$ is the collection of all orthogonal matrices in $\reals^{k}$. Let $O_t= \arg\min_{O\in O(k)}\|Z_t-Z^*O\|_F$, $\Delta_{Z_t}=Z_t-Z^*O_t$, 
and $\Delta_{\Theta_t}=\Theta_t-\Theta^*$. 

For theoretical justification, in Algorithm~\textcolor{cardinal}{S1}, we further assume projection onto the constraint sets $\mathcal{C}_Z=\{Z\in\reals^{n\times k}, J_nZ=Z, \max_{1\leq i\leq n} \|Z_{i*}\|^2\leq M_1/2\}$ and $\mathcal{C}_\alpha=\{\alpha\in \reals^n, 2\|\alpha\|_{\max} \leq M_1/2 \}$ at each iteration. 
The following proposition establishes the high probability error bounds for estimating both the latent position matrix $Z$ and the logit-transformed probability matrix $\Theta$.

\begin{proposition}
\label{prop: separate est z error rate}
	1) the initializers $\alpha_0, Z_0$ in Algorithm~\textcolor{cardinal}{S1} satisfy $\|Z^*\|_{op}^2\|\Delta_{Z_0}\|_F^2 + \|\Delta_{\alpha_0} 1_n^\top\|_F^2 \leq c_0e^{-2M_1}\|Z^*\|_{op}^4/\kappa_{Z^*}^4$ for a sufficiently small positive constant $c_0$, where $\kappa_{Z^*}$ is the condition number of $Z^*$; and 2) $\|Z^*\|_{op}^2\geq C_1 \kappa_{Z^*}^2\sqrt{n}e^{M_1-M_2/2} \max\{\sqrt{\tau k} e^{M_1}, 1\}$ for a sufficiently large constant $C_1$.
	Then there exist positive constants $\rho, c_1,$ and $C$ uniformly over $\mathcal{F}_{\theta}(n, k, M_1, M_2)$ such that, with probability at least $1-n^{-c_1}$, 
we have
	\[\|Z^*\|_{op}^2\|\Delta_{Z_T}\|_F^2, \ \|\Delta_{\Theta_T}\|_F^2\leq C\kappa_{Z^*}^2e^{2M_1}nk\cdot \max\{e^{-M_2}, \frac{\log n}{n}\},\]
	for some $T\leq \log(\frac{M_1^2}{\kappa_{Z^*}^2 e^{4M_1-M_2}}\frac{n}{k^3})/\log (1-\frac{\tau}{e^{M_1}\kappa_{Z^*}^2}\rho)^{-1}$.
\end{proposition}

\vpara{Error bound for Algorithm~\textcolor{cardinal}{S2}} 
Let $v_t$ be the updated parameters at the $t$-th iteration in Algorithm~\textcolor{cardinal}{S2} and $\eta_t=v_t{v_t}^\top$. 
Similarly, as the latent polar variables $v\in \reals^{n}$ are identifiable up to a sign, we define the distance between two latent vectors $v_1$ and $v_2$ as 
$dist(v_1, v_2) = \min_{\kappa\in \{-1,1\}}\|v_1-\kappa v_2\|$. 
Let $\kappa_t= \arg\min_{\kappa\in \{-1,1\}}\|v_t-\kappa v^*\|$ and $\Delta_{v_t}=v_t-\kappa_t v^*$, and further let $\Delta_{\eta_t}=\eta_t-\eta^*$. 

 Although the error bound presented below does not rely on a specific generating process of edges such as in model~(\ref{eq: model step 1}) and the parameter space $\mathcal{F}_{\theta}(n, k, M_1, M_2)$ in (\ref{eq: para space 1}), it depends on the lower bound of the  probability of observing an edge. For notation consistency, we use $M_1$ to denote the lower bound of the logit-transformed probability matrix, i.e., $\Theta_{ij} \geq -M_1$ for $1\leq i, j\leq n$. 
Similarly, for theoretical justification, we constrain $v$ to be in the set $\mathcal{C}_v=\{v\in\reals^{n}, \|v\|_{\max}^2 \leq M_3 \}$ at each iteration in Algorithm~\textcolor{cardinal}{S2}. 
The following theorem establishes the high probability error bounds for estimating the latent polar variables $v$ and the logit-transformed probability matrix  $\eta$.

\begin{theorem}
\label{thm: separate est v error rate}
	Set the step size as $\tau_v=\tau/\|v_0\|^2$ for any $\tau\leq c$, where $c>0$ is a universal constant. Suppose  
	1) the initializer $v_0$ in Algorithm~\textcolor{cardinal}{S2} satisfies $\|\Delta_{v_0}\|\leq c_0e^{-(M_1+M_3)/2}\|v^*\|$ for a sufficiently small positive constant $c_0$; 
	and 2) $\|v^*\|^2\geq C_1 \sqrt{n} e^{M_1+M_3}\max\{\sqrt{\tau e^{M_1+M_3}}, 1\}$ for a sufficiently large constant $C_1$. Then there exist positive constants $\rho, c_1,$ and $C$ uniformly over $\mathcal{F}_{\eta}(n,M_3)$ and $M_1$ such that, with probability at least $1- n^{-c_1}$, we have 
		\[\|v^*\|^2\|\Delta_{v_T}\|^2, \  \|\Delta_{\eta_T}\|_F^2 \leq Ce^{2(M_1+M_3)}n,\]
		for some $T\leq \log(\frac{M_3^2 }{e^{3(M_1+M_3)}}n)/\log(1-\frac{\tau}{e^{M_1+M_3}}\rho)^{-1}$.
\end{theorem}

Theorem~\ref{thm: separate est v error rate} implies that the mean square error $\|\Delta_{\eta_T}\|_F^2 / n^2$ is of order $\mc{O}(1/n)$, which coincides with the existing error rate for one-bit rank-$1$ matrix completion problems~\parencite{davenport20141, chen2015fast}, while our result can be viewed as their extension to the case where entries of the one-bit matrix are randomly observed with non-uniform probabilities. In particular, for the more general non-uniform case, the key ingredient in our proof is to derive a lower bound of the sampling operator $|A| \in \{0,1\}^{n\times n}$. We prove that the sampling operator $|A|$ has a positive lower bound, i.e., $\||A|\circ \eta\|_F \geq c\|\eta\|_F$ with some $c>0$, as long as $\eta$ belongs to a specific data-dependent set. This positive lower bound enables us to extend the proof in \textcite{ma2020universal} when establishing iterative improvements. 
The proof of Theorem~\ref{thm: separate est v error rate} is given in the Supplemental Material.

\begin{remark}
Note that the error bounds for $v$ and $\eta$ in Theorem~\ref{thm: separate est v error rate} hold regardless of the concrete form of model~(\ref{eq: model step 1}). Therefore, the above results still hold even if model~(\ref{eq: model step 1}) is mis-specified. But, it still depends on the probability of observing edges. 
\new{Recall that the probability of forming edges are bounded by $e^{-M_1} \asymp 1/(1+e^{M_1}) \le P_{ij} \le 1/(1+e^{M_2}) \asymp e^{-M_2}$.
For $M_1 \asymp M_2$, the network density is in the order of $\rho=e^{-M_1}$. Theorem~\ref{thm: separate est v error rate} implies that the error bound decreases as the network density $\rho$ grows. Specifically, the impact of network density lies in the multiplier $1/\rho^2$. 
Intuitively, a sparse network, with fewer observed edges, leads to larger estimation errors for $v$ and $\eta$ due to the lack~of~sign observations. 
Similarly, Proposition~\ref{prop: separate est z error rate} implies that the network density affects the error bound of $Z$ through the multiplier $\max\{\rho, \log n / n\} / \rho^2$. For networks with modest density ($\rho \geq \log n /n$), the multiplier is $1/\rho$. For sparser networks where $\rho \leq \log n / n$, the multiplier adjusts to $\log n /(n\rho^2)$.
}
\end{remark}

\begin{remark}
The assumptions in both Proposition~\ref{prop: separate est z error rate} and Theorem~\ref{thm: separate est v error rate} require relatively good initializations of $(\alpha, Z, v)$. We note that the conditions for $\alpha_0$ and $Z_0$ can be achieved with theoretical justification by the universal-singular-value-thresholding \parencite{chatterjee2015} based initialization algorithm proposed in \textcite{ma2020universal}. We further extend this algorithm to initialize $v_0$. Based on our simulation studies (see the Supplemental Material), we find that simple random initialization also achieves similar estimation errors after the algorithm converges while requiring more iterations for algorithm convergence. 
\end{remark}

\subsection{Results for the joint estimation method}
\label{subs: theory for joint}
We first present the convergence guarantee and the error bound for the estimators obtained by Algorithm~\textcolor{cardinal}{S5}. Then we further investigate how the joint estimation method could improve the estimation of $Z$ on top of the separate estimation.

Under the joint inner-product model, we redefine the parameter space as 
\begin{align*}
	\mathcal{F}(& n, k, M_1, M_2, M_3)= \Big\{\alpha, v \in \reals^n, Z \in \reals^{n\times k}, \Theta, \eta \in \reals^{n \times n}  \ \big|  \\
	& \ \ \ \ \  \Theta=\alpha 1_n ^\top+1_n\alpha^\top+ZZ^\top, J_nZ=Z, \eta=vv^\top, v =Zw+\gamma1_n, \\
	& \ \ \ \  \max_{1\leq i\leq n} \|Z_{i*}\|^2, \|\alpha\|_{\max} \leq \frac{M_1}{2}, \max_{1\leq i\neq j\leq n}\Theta_{ij}\leq -M_2, \|v\|_{\max}^2\leq M_3, \|w\|\leq M, |\gamma|\leq M' \Big\},
\end{align*}
where $k$, $M_1$, $M_2$, and $M_3$ are allowed to change with the network size $n$. 
Let $(\alpha^*, Z^*, v^*)$ be the true parameters, where $v^*= Z^*w^*+\gamma^*1_n$ with some $w^*\in\reals^k$ and $\gamma^*\in\reals$.  

\vpara{Error bound for Algorithm~\textcolor{cardinal}{S5}}
Let $(\alpha_t, Z_t, v_t)$ be the updated parameters at the $t$-th iteration in Algorithm~\textcolor{cardinal}{S5}. We assume the projection onto the same constraint sets $\mc{C}_{\alpha}$, $\mc{C}_Z$, and $\mc{C}_v$ at the end of each iteration as those for Algorithms~\textcolor{cardinal}{S1} and~\textcolor{cardinal}{S2}. The following theorem first guarantees that the error of iterates $\{(\alpha_t, Z_t)\}_{t\ge 0}$ converges up to a statistical error and then gives the high probability error bounds for the estimators of $Z$ and $\Theta$. 

\begin{theorem}
\label{thm: joint error rate}
Set the step sizes as $\tau_Z = r_0 \tau/\|Z_0\|^2_{op}$, $\tau_\alpha = \tau / (2n)$, and the weight $\lambda=\tilde{\lambda}r_0/e^{M_1}\kappa^2_{Z^*}$ with $r_0=\min\{1, \|Z_0\|^2_{op}/\|v_0\|^2\}$ for any $\tau \le c_\tau$, $\tilde{\lambda}\le c_{\lambda}$, where $c_\tau$ and $c_\lambda$ are universal constants. Let $\zeta_n=\max\{\||A|-P\|_{op}, 1\}$ and $\varphi_n=\max\{\||A|\circ ((1+A)/2-Q)\|_{op}, 1\}$. Denote the error metric for iterates as $\tilde{e}_{Z,t} = \|\Delta_{Z_{t}}\|_F^2 \|Z_0\|_{op}^2+\|\Delta_{\alpha_t} 1_n^\top\|_F^2$. 
Suppose the initializers $\alpha_0, Z_0$ in Algorithm~\textcolor{cardinal}{S5} satisfy $\tilde{e}_{Z,0} \leq c_0e^{-2M_1-3M_3}\|Z^*\|_{op}^4/\kappa_{Z^*}^4$ for a sufficiently small positive constant $c_0$, where $\kappa_{Z^*}$ is the condition number of $Z^*$.~Then,~we~have
\begin{enumerate}
	\item (Deterministic bounds for iterative errors) If $\|Z^*\|_{op}^2\ge C_0 e^{M_1} \kappa_{Z^*}^2 \zeta_n \max\{\sqrt{\tau k}e^{M_1+3M_3/2}\kappa_{Z^*}, 1\}$ and $\|v^*\|^2\geq C_0 e^{M_1+M_3}\varphi_n \max\{\sqrt{\tau}e^{M_1/2+M_3},1\}$  for a sufficiently large constant $C_0$, then there exist universal positive constants $\rho_1$, $\rho_2$, $C'$, and $C''$ such that for all $t\ge 0$
		\begin{align*}
			\tilde{e}_{Z, t+1}\le & \left(1-\frac{r_0 \tau \rho_1}{e^{M_1}\kappa^2_{Z^*}} \right)\tilde{e}_{Z,t} -\lambda \frac{r_0 \tau  \rho_2}{e^{M_3} } \min\{\||A|\circ \Delta_{\eta_t}\|_F^2, e^{-M_1}\|\Delta_{\eta_t} \|_F^2\}  \\
			&\ \  +  r_0\tau C'e^{M_1}\zeta_n^2 k  + \lambda r_0\tau  C''  e^{M_1+M_3}\varphi_n^2. 
		\end{align*}
	\item (High-probability bounds) Suppose $\|Z^*\|_{op}^2\ge C_0 e^{M_1-M_2/2} \kappa_{Z^*}^2 \sqrt{n} \max\{\sqrt{\tau k}e^{M_1+3M_3/2}\kappa_{Z^*}, 1\}$ and $\|v^*\|^2 \geq C_0 e^{M_1+M_3}\sqrt{n} \max\{\sqrt{\tau}e^{M_1/2+M_3},1\}$  for a sufficiently large constant $C_0$. Then there exist positive constants $\rho_1$, $c$, and $C$ uniformly over $\mathcal{F}(n,k, M_1, M_2, M_3)$ such that, with probability at least $1-n^{-c}$, we have
	\[\|Z^*\|_{op}^2\|\Delta_{Z_T}\|_F^2, \ \|\Delta_{\Theta_T}\|_F^2\leq C \kappa_{Z^*}^2 e^{2M_1} nk \cdot \max \{ e^{-M_2}, \frac{\log n}{n},  e^{M_3-M_1}\frac{1}{\kappa_{Z^*}^2 k}\},\]
	for some $T\leq \log(\frac{M_1^2}{\kappa_{Z^*}^2 e^{4M_1+3M_3-M_2}}\frac{n}{k^3})/\log (1-\frac{r_0\tau \rho_1}{e^{M_1}\kappa_{Z^*}^2})^{-1}$.
\end{enumerate}
\end{theorem}

The first part of Theorem~\ref{thm: joint error rate} indicates that, compared to the separate estimation method, the joint method involving the gradient of the sign likelihood leads to an extra improvement on the error bound of iterates, which depends on $\Delta_{\eta_t}$, while introducing another statistical error term $\varphi_n$. As a result, in the second part, the high probability error bounds depend on the maximum of three terms, among which the first two are the same as in Proposition~\ref{prop: separate est z error rate} and the third is resulted from~$\varphi_n$. When $M_3 \le M_1-M_2+\log(\kappa_{Z^*}^2 k)$, the maximum multiplier reduces to the one in Proposition~\ref{prop: separate est z error rate}. Overall, the error bounds of $Z$ and $\Theta$ for the joint estimation method are still in the order $\mc{O}(nk)$, which is the same as that for the separate estimation method. In addition, the following corollary gives the error bounds of $v_T$ and $\eta_T$ obtained from the line~5 in Algorithm~\textcolor{cardinal}{S5}.

\begin{corollary}
\label{corollary: error bound v} For $v_t=Z_tw_t+\gamma_t1_n$ with $(w_{t}, \gamma_{t})=\argmin_{w\in \reals^k, \gamma \in \reals} \mc{L}_\lambda(\alpha_{t}, Z_{t}, w, \gamma)$, we have 
$\|\Delta_{\eta_t}\|_F \le   16 e^{M_1+M_3}\max\{ \zeta_n, \varphi_n\}+e^{M_1/2+M_3}(2+\|\Delta_{Z_t}\|_F \|w^*\|)\|v^*\|\|\Delta_{Z_t}\|_F \|w^*\|$ for $t \ge 0$. 
Suppose the conditions for high probability bounds in Theorem~\ref{thm: joint error rate} hold, then there exist positive constants $\rho_1$, $c$, and $C$ uniformly over $\mathcal{F}(n,k, M_1, M_2, M_3)$ such that, with probability at least $1-n^{-c}$, we have
	\[\|v^*\|^2 \|\Delta_{v_T}\|^2, \|\Delta_{\eta_T}\|^2 \leq  C e^{3M_1+2M_3}nk \cdot \max\{\frac{e^{M_3-M_1}}{k}, \kappa_{Z^*}^2 \max \{ e^{-M_2}, \frac{\log n}{n}\}\},\]
	for some  $T\leq \log(\frac{M_1^2}{\kappa_{Z^*}^2 e^{4M_1+3M_3-M_2}}\frac{n}{k^3})/\log (1-\frac{r_0\tau \rho_1}{e^{M_1}\kappa_{Z^*}^2})^{-1}$.
\end{corollary}

In particular, the deterministic error bound for $\Delta_{\eta_t}$ consists of the statistical error term $\max\{\zeta_n, \varphi_n\}$ and the estimation error of $Z_t$, and with high probability $\|\Delta_{\eta_T}\|^2$ is dominated by the estimation error of $Z_t$ and thus is also in the order $\mc{O}(nk)$.

\vpara{One-step improvement}
Although Theorem~\ref{thm: joint error rate} guarantees the convergence of Algorithm~\textcolor{cardinal}{S5} up to certain statistical errors, the achieved error rate is in the same order as that for the separate estimation method.
To further investigate how exploiting extra structural information in the joint inner-product model would help estimate the latent variables, we consider the estimation error moving against the gradient one step from the estimators obtained by the separate method below.

Suppose we are given estimators $(\bar{\alpha},\bar{Z})$ of latent variables obtained from the separate estimation Algorithm~\textcolor{cardinal}{S1} and an estimator $\bar{v} = \bar{Z}\bar{w}+\bar{\gamma}1_n $. 
Then we update the estimator of $Z$ by one step through Algorithm~\textcolor{cardinal}{S5} as below:
\begin{equation}
	\label{eq: update z}
	\hat{Z}= \bar{Z}-2\tau_z(1-\lambda) (\sigma(\bar{\Theta})-|A|)\bar{Z}- 2\tau_z\lambda\big(|A|\circ \sigma(\bar{\eta})- B\big)\bar{v} \bar{w}^\top,
\end{equation}
where $\bar{\Theta} = \bar{\alpha} 1_n ^\top+1_n{\bar{\alpha}}^\top+\bar{Z}\bar{Z}^\top$, $\bar{\eta} = \bar{v}\bar{v}^\top$, and $B =|A|\circ(A+1)/2$. 
Note that $[B]_{ij}=|A_{ij}|b_{ij}$ with $b_{ij}$ independently following Bernoulli$(\sigma(\eta_{ij}^*))$ conditional on $|A|$.
The following proposition provides insights on under what scenarios the one-step update in the joint estimation method could lead to better estimates of the latent position vectors $Z$. In below, for ease of derivation, we consider the parameter space $\mathcal{F}(n, k, M_1, M_2, M_3)$ with fixed $M_i$ ($i=1,2,3$) and $k$.  
\begin{proposition}
\label{prop: improvement on Z}
Given the estimators $(\bar{\alpha}, \bar{Z})$ obtained from Algorithm~\textcolor{cardinal}{S1} and the estimators~$(\bar{w}, \bar{\gamma})$ that are independent of $B$ conditional on $|A|$ and satisfy $\|\bar{w} - w^*\|^2 + \|\bar{\gamma}-\gamma^*\|^2=\mathcal{O}(1/n)$. 
We update $\bar{Z}$ for one step by (\ref{eq: update z}) and obtain $\hat{Z}$. 
Suppose the conditions in Proposition~\ref{prop: separate est z error rate} hold, and the singular values of the sample covariance ${Z^*}^\top Z^* / n$ are of constant order. 
Then there exists an optimal $\lambda_{opt}$ that minimizes $\mathbb{E}\|\hat{Z}-Z^*\|_F^2$. Furthermore, if $\lambda_{opt}>0$, we have $\mathbb{E}\|\Delta_{\hat{Z}}\|_F^2 < \mathbb{E} \|\Delta_{\bar{Z}}\|_F^2$ for any $\lambda \in (0, 2\lambda_{opt})$, and the improvement $\mathbb{E}\|\Delta_{\bar{Z}}\|_F^2 - \mathbb{E} \|\Delta_{\hat{Z}}\|_F^2$ with $\lambda_{opt}$ is at least 
\[ \frac{ \big\||A|\circ\xi \circ T_1\big\|_F^2 \left(  \big\||A|\circ\xi \circ T_1\big\|_F - \big\||A|\circ\xi \circ T_2 \big\|_F-\big\||A|\circ\xi \circ T_3 \big\|_F \right)^2}{16 \left(\big\||A|\circ\xi \circ \xi \circ(T_1 + T_2 - T_3) \big\|_{op}^2 + \mathbb{E}\big\| B-|A|\circ \sigma(\eta^*) \big\|_{op}^2 / (\|\bar{Z}\|_{op}^2\|\bar{w}\|^4) \right)},\]
where $T_i$'s are given in (\textcolor{cardinal}{S42})-(\textcolor{cardinal}{S44}) respectively for $i=1,2,3$ in the Supplemental Material with  $\|T_1\|_F =\mc{O}(1)$, $\|T_2\|_F =\mc{O}(1)/\|w^*\|$, and $\|T_3\|_F =\mc{O}(1/\sqrt{n})$, and $\xi$ is an element-wise positive constant matrix.
Here $\mathbb{E}$ represents the conditional expectation of $B$ given $|A|$.
\end{proposition}
We provide the expression of the optimal $\lambda_{opt}$ that minimizes $\mathbb{E}\|\hat{Z}-Z^*\|_F^2$, the proof of Proposition~\ref{prop: improvement on Z}, and discuss when the conditional independence assumption and the prerequisite error rate of $(\bar{w}, \bar{\gamma})$ in Proposition~\ref{prop: improvement on Z} hold in the Supplemental Material. Since a positive $\lambda_{opt}$ implies a strict decrease in the mean square error of $Z$ after one-step update, we further investigate in which case $\lambda_{opt}$ tends to be positive. In particular, $\lambda_{opt}>0$ if and only if $\big\||A|\circ\xi \circ T_1\big\|_F - \big\||A|\circ\xi \circ T_2 \big\|_F-\big\||A|\circ\xi \circ T_3 \big\|_F $ is strictly positive. Our analysis on the upper bounds of the three terms suggests that the first two terms are the dominating terms and a larger $\|w^*\|$ more likely results in a positive $\lambda_{opt}$. Therefore, when the signal from the edge sign distribution is strong, incorporating information from observed signs in the joint estimation method is useful for improving the estimation of $Z$. Moreover, the magnitude of improvement also depends on the levels of the signal and the noise in the sign distribution. Specifically, as  $\|w^*\|\asymp \|\bar{w}\|$ increases, the difference between the upper bounds of the two dominating terms in the numerator increases while the upper bound of the denominator decreases, therefore overall the improvement is likely to increase. This implies that larger signals in the edge signs would lead to greater improvement in estimating $Z$. On the other hand, we find that the improvement decreases when the noise $\mathbb{E}\big\| B-|A|\circ \sigma(\eta^*) \big\|_{op}^2 / \|\bar{Z}\|_{op}^2$ in the edge sign distribution increases in the denominator.

\section{Simulation Studies}
\label{sec: simu}

In this section, we conduct  simulation studies to investigate how estimation errors of the proposed methods depend on: 1) the network size and the dimension of latent position vectors; 2) the network density; and 3) the proportion of positive edges.  

\vpara{Estimation methods} We compare three estimation methods. In addition to the separate estimation method and the joint estimation method introduced in Section~\ref{sec: est}, we further add an intermediate method, \textit{one-step-joint} estimation, to illustrate the one-step improvement discussed in Proposition~\ref{prop: improvement on Z}. 
Specifically, given $\bar{Z}$ and $\tilde{v}$ obtained from Algorithms~\textcolor{cardinal}{S1} and \textcolor{cardinal}{S2} respectively, we compute the one-step-joint estimators $(J_n\hat{Z}, \bar{v})$ by first updating $\bar{v} = \bar{Z}\bar{w}+\bar{\gamma}1_n$ with $\bar{w}, \bar{\gamma}=\argmin_{w\in\reals^k,\gamma\in\reals}\|\tilde{v}-\bar{Z}w-\gamma1_n\|$ and then obtaining $\hat{Z}$ by plugging $(\bar{Z}, \bar{v})$ into (\ref{eq: update z}). We set $\lambda=1/2$, so that the joint estimation is the same as the maximum likelihood estimation.

\vpara{Simulation settings} For a given network size $n$ and a latent position vector dimension $k$, we set the model parameters as follows.
We first generate the latent positions $Z_{ij}\overset{\text{iid}}{\sim} \mathcal{N}(0,1) $ from the standard normal distribution, for $1\leq i\leq n, 1\leq j\leq k$. By centering columns of $Z$, we get $Z^* = J_nZ$, where~$J_n=I_n-\frac{1}{n} 1_n 1_n^\top$. We further normalize $Z^*$ element-wise such that $\|Z^*{Z^*}^\top\|_F=n$. Next, we generate the node degree heterogeneity parameters $\alpha_i^*= - \alpha_i / \sum_{i=1}^n\alpha_i$, where $\alpha_i\overset{\text{iid}}{\sim}U(1,3)$ is uniformly distributed for $1\leq i\leq n$. Finally, we set $w^*=1/\sqrt{k}\cdot 1_k$, $\gamma^*=0$, and $v^*_i={w^*}^\top Z^*_i +\gamma^*$.

Given the true latent variables $Z^*, \alpha^*$, and $v^*$, we randomly generate 20 replications of the signed adjacency matrix following (\ref{eq: model step 1}) and (\ref{eq: model step 2}), and fit models by three estimation methods. 
For each method, we measure the relative errors for $Z, v, \Theta$, and $\eta$. Due to the identifiability conditions in Proposition~\ref{prop: iden for linear}, we define the relative error for $Z$ as $\|\hat{Z}-Z^*\bar{Q}\|_F/\|Z^*\|_F$, where $\bar{Q} = \argmin_{Q\in O(k)}\|\hat{Z}-Z^*Q\|_F$ and $O(k)$ is the collection of all orthogonal matrices in $\reals^{k}$. 
We define the error for $v$ as $\|\hat{v} - \bar{\kappa} v^*\|/\|v^*\|$, where $\bar{\kappa}=\argmin_{\kappa\in \{1,-1\}}\|\hat{v} - \kappa v^*\|$. The relative errors for $\Theta$ and $\eta$ are defined as $\|\hat{\Theta}-\Theta^* \|_F / \|\Theta^*\|_F$ and $\|\hat{\eta}-\eta^* \|_F / \|\eta^*\|_F$ respectively, where $\hat{\Theta}=\hat{\alpha} 1_n ^\top+1_n\hat{\alpha}^\top+\hat{Z}\hat{Z}^\top$ and $\hat{\eta}=\hat{v}\hat{v}^\top$. 

\begin{figure}[!tb]
	\centering
	\includegraphics[width=0.79\linewidth]{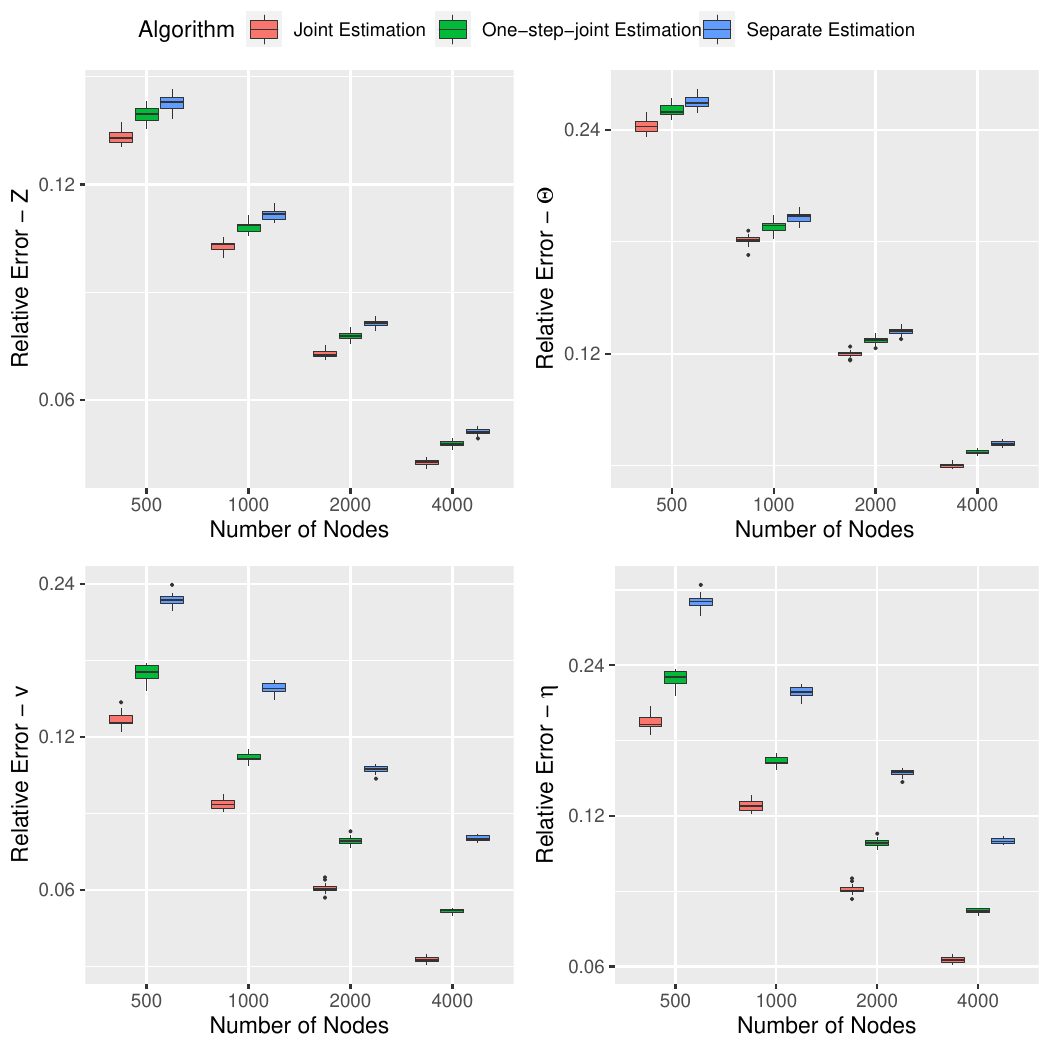}
	\caption{Log-log plots of relative errors with respect to the network size $n$. The dimension of the latent position vector is fixed as $k=2$.}
	\label{fig:sim-size}
\end{figure}

\subsection{Varying the network size and the dimension of the latent~space}
In Figure~\ref{fig:sim-size}, we summarize how estimation errors vary with different network sizes. We fix $k=2$ and vary $n\in\{500, 1000, 2000, 4000\}$. We can see that, for a fixed dimension of the latent space, the relative errors of all three estimation methods decrease in the rate of $1/\sqrt{n}$ as the network size $n$ grows, which align well with the theoretical error rates given in Section~\ref{sec: thm}. 
Next, compared to the separate estimation method, the joint estimation method consistently achieves smaller estimation errors on all four quantities of interest across different network sizes. 
In addition, the one-step-joint estimation that simply updates estimates by one-step gradient descent is able to reduce the estimation errors compared to the separate estimation method.

We further summarize how estimation errors of $Z$ and $\Theta$ vary with different dimensions of the latent position vector in Figure~\textcolor{cardinal}{S3} in the Supplemental Material for the sake of space. We fix $n=2000$ and vary $k\in\{2, 4, 8\}$. 
We find that, for a fixed network size, the relative errors increase in the rate of $\sqrt{k}$ as the dimension of latent position vector $k$ grows. This also agrees well with our theoretical results.
The relative trend among the three estimation methods for different $k$ is similar as that in Figure~\ref{fig:sim-size}, where the joint estimation method is consistently the best.

\vspace{-5mm}
\subsection{Varying the network density}
\vspace{-3mm}
We investigate how estimation errors for three estimation methods vary with the network density. To this end, we generate the node degree heterogeneity parameters $\alpha_i^* = -\bar{\alpha}- \alpha_i / \sum_{i=1}^n\alpha_i$ where $\alpha_i\overset{\text{iid}}{\sim}U(1,3)$ is uniformly distributed for $1\leq i\leq n$. We fix $n=2000$ and $k=4$, and vary $\bar{\alpha}\in \{0, 0.25, 0.5, 0.75, 1, 1.25\}$, which leads to the network density ranging from $0.1$ to $0.5$. 

Figure~\textcolor{cardinal}{S4} in the Supplemental Material summarizes the relative estimation errors of $Z$ and $v$ over $20$ replications under different network densities. We can see that both estimation errors of $Z$ and $v$ for all three estimation methods decrease as the network gets denser, and the joint estimation method  achieves lower estimation errors than the other two methods consistently across various network densities. In particular, when the network is dense, the improvement in estimating $Z$ from the joint estimation against the separate estimation increases, which is expected because in joint estimation, the observed edges' signs are also useful for inferring $Z$ and denser networks provide more information. In addition, regarding the estimation error of $v$, the joint and one-step-joint estimation methods that use the additional structural information between $Z$ and $v$ perform more stably than the separate estimation method as the network gets sparser. 

\begin{figure}
	\centering
	\includegraphics[width=0.79\linewidth]{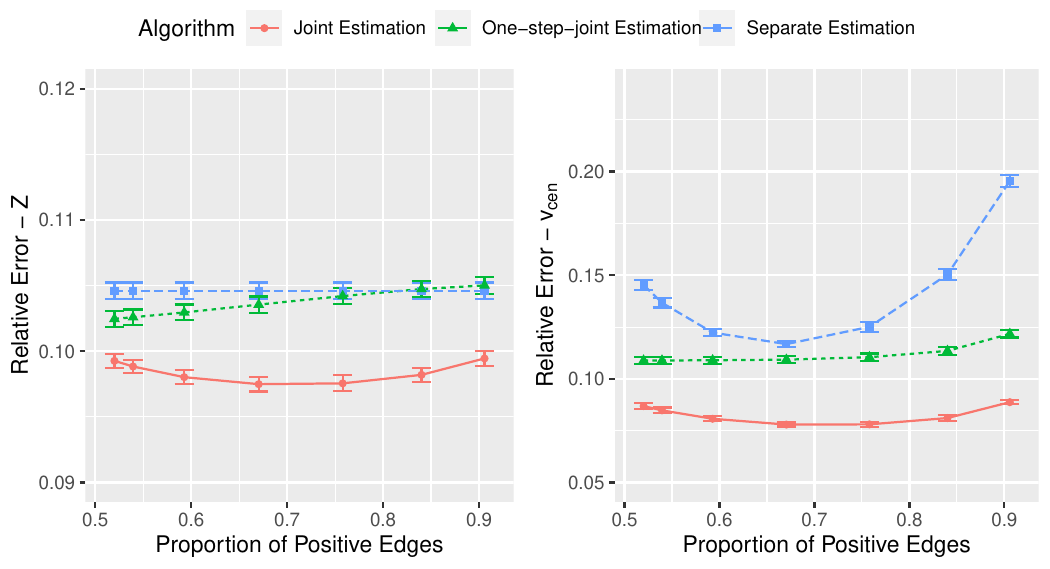}
	\caption{Relative errors with respect to the proportion of positive edges. The network size is fixed as $n=2000$ and the dimension of the latent position vector is fixed as $k=4$.}
	\label{fig: sign ratio}
\end{figure}

\subsection{Varying the proportion of positive edges}
We also investigate the effect of the proportion of positive edges on three estimation methods. For this purpose, we change the simulation setting. Specifically, we fix $n=2000$ and $k=4$, and vary $\gamma^* \in \{0,0.3,0.6,0.9,1.2,1.5,1.8\}$, which results in the proportion of positive edges ranging from $0.52$ to $0.91$. To eliminate the artificial effect resulting from varying $\gamma^*$ when evaluating the estimation error of $v$, we focus on the estimation error of the centered $v$, i.e., $v_{cen}=J_n v$. We define the relative error for $v_{cen}$ as $\|J_n\hat{v} - \bar{\kappa} J_nv^*\|/\|J_nv^*\|=\|J_n\hat{v} - \bar{\kappa} Z^* w^*\|/\|Z^* w^*\|$, where $\bar{\kappa}=\argmin_{\kappa\in \{1,-1\}}\|J_n\hat{v} - J_n\kappa v^*\|$.  

Figure~\ref{fig: sign ratio} summarizes the relative estimation errors of $Z$ and $v_{cen}$ over $20$ replications under different proportions of positive edges. Overall, the joint estimation method performs consistently the best among three methods and is robust across different sign distributions. We note that the estimation error for $Z$ for the separate method does not change, because the generated absolute adjacency matrix $|A|$ does not change when varying $\gamma^*$ and thereby the separate estimates $\hat{Z}$ stay the same. \new{We also observe that the optimal performance of the separate method for estimating $v$ is not achieved around 50\% positive signs. Intuitively, in extreme cases where $v\approx 0$, the negative and positive signs become indistinguishable, with the proportion of positive edges close to 50\%. This makes it challenging to estimate $v$. The optimal proportion minimizing the estimation error may vary case by case. We investigate an SBM example in the Supplemental Material for illustration.}

\section{International Relation Data}
\label{sec: real data}

In this section, we apply the proposed method to an international relation data, i.e. the Correlates of War (COW)~\parencite{izmirlioglu2017correlates}, to demonstrate how the proposed method can be used to make informative interpretation and visualization of signed networks. The COW dataset records various types of international relations among countries, such as wars, alliances, and militarized interstate disputes. Similarly as \textcite{PhysRevE.99.012320}, we construct a signed network of countries, where the positive edges represent alliance relationships, and the negative edges represent the existence of militarized disputes between countries. We take the snapshot of the records during World War II (WWII), i.e., from 1939 to 1945. In particular, if two countries were involved in both alliances and militarized disputes, we set the sign of their edge to positive if the number of years of alliances is larger than that of the militarized disputes, and we set the sign to negative otherwise. According to the COW records, there are 68 countries that were involved in alliances or militarized disputes during WWII, and the resulted signed network contains 566 positive edges and 519 negative edges. \new{The total number of balanced triangles is 2,148, outnumbering the unbalanced triangles (613) by over threefold. To further assess the balance property, we applied a stratified permutation test~\parencite{doi:10.1080/01621459.2020.1764850}. Under the null hypothesis, the sign of an edge is independent of its position on the graph, which is considered as balance-free. The p-value of the stratified permutation test is less than $0.001$, which indicates strong evidence of the balance property in this international relation network.}

\begin{figure}[!tb]
    \centering
    \includegraphics[width=0.86\textwidth]{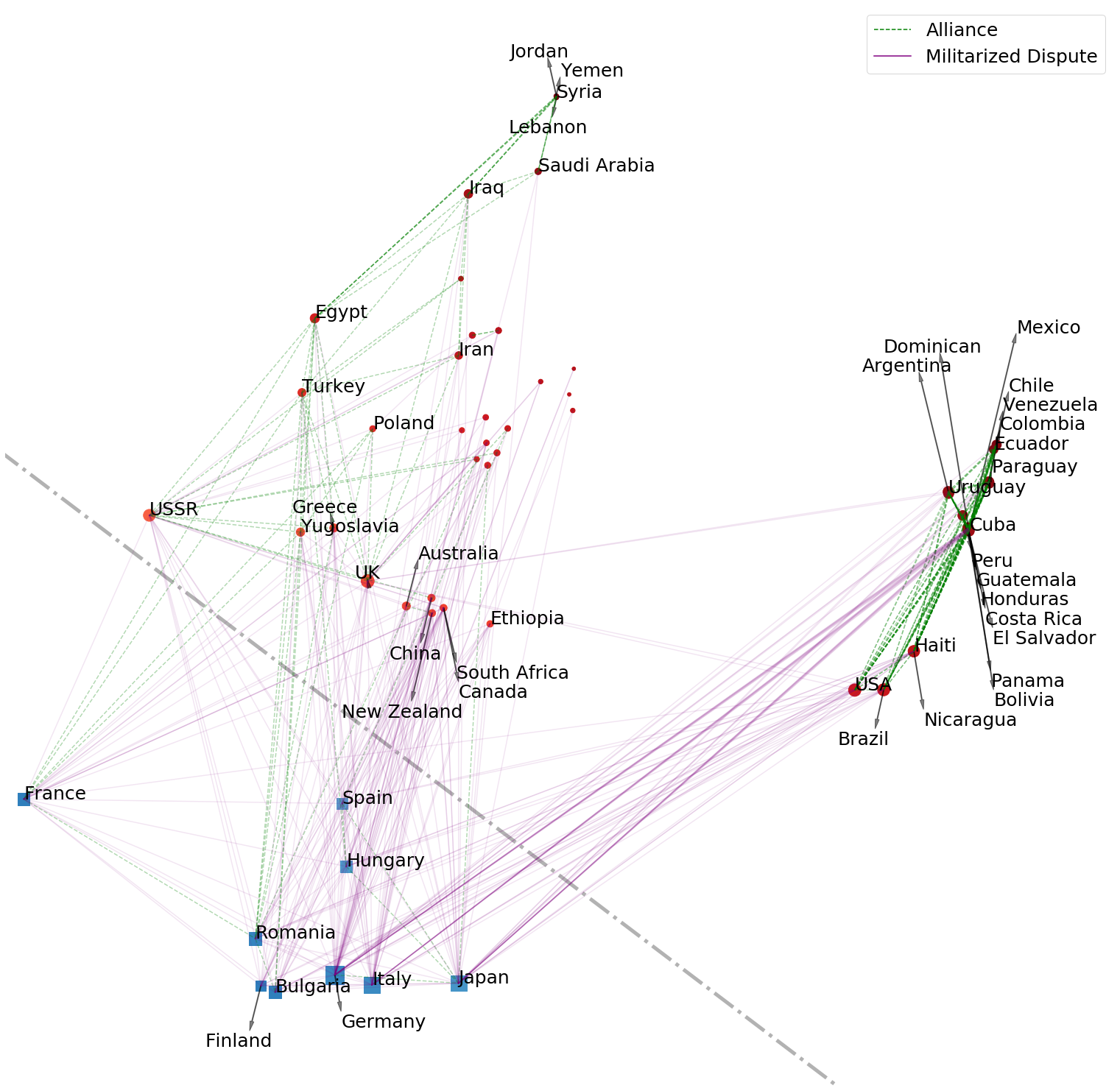}
    \caption{Visualization of the model fitted by the joint estimation method on the COW dataset. The nodes are countries involved in WWII. The green dashed lines represent positive edges (alliance) and the purple solid lines represent negative edges. The node sizes are determined by the estimated node degree heterogeneity parameters $\hat{\alpha}$. The node colors and shapes are determined by the estimated latent polar variables $\hat{v}$: for each node $i$, the node is a red circle if $\hat{v}_i > 0$ and a blue square otherwise; and the larger absolute magnitude of $\hat{v}_i$, the darker the color. The grey dash-dotted line represents the linear boundary between the red and blue nodes. To make the visualization easier, we only labeled countries with degree greater than 5.}
    \label{fig:joint_ww2}
\end{figure}

To incorporate the balance theory, we fit the joint inner-product model to the COW dataset. 
We set the dimension of the latent position vectors as $k=2$, such that the estimated $\hat{Z}$ can be directly visualized on a 2-dimensional plane. 
The model fitted by the joint estimation method is visualized in Figure~\ref{fig:joint_ww2}.
In the figure, each node represents a country and their coordinates are given by $\hat{Z}$. The size of each node is determined by the estimated degree heterogeneity parameter $\hat{\alpha}$, with larger nodes corresponding to larger $\hat{\alpha}_i$ values. The color and the shape of each node $i$ distinguish the estimated latent polar variable $\hat{v}_i$. Specifically, if $\hat{v}_i > 0$, the node is visualized as a red circular point; and if $\hat{v}_i < 0$, the node is visualized as a blue square point. For both the red and the blue points, the larger the absolute magnitude $|\hat{v}_i|$, the darker the color. The sign of each edge is also indicated in the figure, with dashed green being positive and solid purple being negative.

Figure~\ref{fig:joint_ww2} shows that the proposed model fitted by the joint estimation method is able to capture important information in the signed network of countries during WWII. First, in terms of the estimated node degree heterogeneity $\hat{\alpha}$, the top 11 countries are Germany, Italy, Japan, the United Kingdom (UK), Romania, the United States (USA), Brazil, Bulgaria, Hungary, France, the Soviet Union (USSR). In particular, UK, USA, USSR were the 3 leading countries of the Allies of WWII. France played important and complicated roles in both the Allies and the Axis. Brazil was the only South American country that actively participated in WWII. All other countries were members of the Axis. Since members of the Axis were more active than those of the Allies on average, it is reasonable that even small members have high values of $\hat{\alpha}$. Second, regarding the estimated latent polar variable $\hat{v}$, the model is able to divide the countries into two groups (blue and red) that mostly align with the division between the Axis and the Allies. 
As we assume a linear transformation from $Z$ to $v$, the plane (the space of $Z$) can be linearly separated into two areas with the boundary illustrated by the grey dash-dotted line. We can see that edges crossing the boundary are mainly negative (purple), while edges within the same side are mainly positive (green).
Finally, the estimated latent position vectors $\hat{Z}$ also capture various other interesting aspects. Within the Allies, $\hat{Z}$ further clusters countries in America together (see the right part of the figure), among which most edges are positive (green).  The Middle East countries form a cluster as well (at the top of the figure), though not as tight as countries in America. It is also interesting to see that France, one of the major Allied powers in WWII, is positioned on the same side of Germany, Italy, and Japan. This is because, after occupied by Germany, France was divided into two political powers, Free France and Vichy France, with the latter collaborating with Germany and fighting against the Allies in several campaigns from 1940 to 1944. \new{We perform additional sensitivity analysis to assess the impact of splitting France into two political entities in the Supplemental Material. }

On the other hand, we also fit the separate inner-product model and visualize the model fitted by the separate estimation method in Figure~\textcolor{cardinal}{S6} in the Supplemental Material for the sake of space. 
We find that the model fitted by the separate estimation method captures useful information from the network but not as interpretable as that by the joint estimation method (see more detailed discussion in the Supplemental Material).

\vspace{-7mm}
\section{Discussion}
\vspace{-3mm}
In this paper, we propose a latent space approach that accommodates the structural balance theory for modeling signed networks. In particular, we introduce a novel notion of population-level balance, which is a natural choice to characterize the structural balance theory when we treat both the edges and their signs as random variables. We develop sufficient conditions for a latent space model to be balanced at the population level, and propose two balanced inner-product models following the conditions. We also provide scalable estimation algorithms with theoretical guarantees. 

\new{
In Section~\ref{sec: models}, we proposed balanced inner-product models that use a one-dimensional polar variable. This design offers a direct and intuitive interpretation, where the sign of the polar variable inherently indicates membership to one of two antagonistic groups. Beyond this, considering multi-dimensional polar variables, i.e., $\eta_{ij}=v_i^\top v_j$ for $v_i \in \reals^d$ with $d>1$, could be more flexible in terms of capturing multiple latent factors in real-world applications. In that case, we have ascertained that our estimation method and high-probability error bounds are still applicable to the multi-dimensional scenario. However, it is also important to note that, when considering multi-dimensional polar variables, the conditions in Theorem 1 no longer hold. In addition, multi-dimensional polar variables have the potential drawback of complicating interpretations.
}

There are a few other directions we may continue to explore in the future. 
First, the joint inner-product model could be extended to have a nonlinear link function $\phi$, which would increase the flexibility of the model. Second, we may generalize the proposed approach to weighted signed networks to better leverage richer edge information available in real-world networks. Finally, it is desirable to extend the latent space approach for undirected networks to directed networks, which can be potentially used to model other interesting social theories, such as the social status theory, for signed networks.





\begingroup
\setstretch{1.7}
\setlength\bibitemsep{0pt}
\AtNextBibliography{\small}
\printbibliography
\endgroup

\end{document}